\documentclass[aps,prd,twocolumn,superscriptaddress]{revtex4-1}
\usepackage[pdftex]{graphicx} % arxiv
\usepackage{color} % arxiv

\usepackage[pdftex,bookmarks=true,bookmarksnumbered=true,colorlinks=true,linkcolor=blue,citecolor=blue,filecolor=blue,urlcolor=blue]{hyperref}  % arxiv

\usepackage{amsfonts}
\usepackage{amsmath}
\usepackage{bm}
\usepackage{mathrsfs}
\usepackage{here}
\newif\ifoneauthor
\oneauthortrue
\newcommand{\unit}[1]{\ {\rm #1}}

\newcommand{\Section}[1]{section \ref{#1}}
\newcommand{\Eq}[1]{Eq. (\ref{#1})}

\newcommand{\Fig}[1]{Figure \ref{#1}}
\newcommand{\HS}{\mathcal{H}_{\text{S}}}
\newcommand{\HV}{\mathcal{H}_{\text{V}}}
\newcommand{\HT}{\mathcal{H}_{\text{T}}}
\newcommand{\HX}{\mathcal{H}_{\text{X}}}
\newcommand{\tc}{t_{\mathrm{c}}}
\newcommand{\phic}{\phi_{\mathrm{c}}}
\newcommand{\dIs}{\{d_I\}^N_{I=1}}

\definecolor{mygreen}{RGB}{0,115,0}

\DeclareMathAlphabet{\mathpzc}{OT1}{pzc}{m}{it}
\usepackage{here} %図を強制的にその位置に出力するためのpackage \begin{figure}[H]でできる。
\definecolor{gray}{gray}{0.4}

\begin{document}

%\preprint{}
%\twocolumn
%\title{Test of gravitational wave polarizations with inspiral waveforms \\of compact binary coalescences}
\title{Pure polarization test of GW170814 and GW170817 \\ using waveforms consistent with modified theories of gravity}



\author{Hiroki Takeda}
\email[]{hiroki.takeda@phys.s.u-tokyo.ac.jp}
\affiliation{Department of Physics, University of Tokyo, Bunkyo, Tokyo 113-0033, Japan}
\author{Soichiro Morisaki}
\affiliation{Department of Physics, University of Wisconsin-Milwaukee, Milwaukee, WI 53201, USA}
\author{Atsushi Nishizawa}
\affiliation{Research Center for the Early Universe (RESCEU), School of Science, University of Tokyo, Bunkyo, Tokyo 113-0033, Japan}

%\thanks{}

%\author{David J. Ottaway}
%\email[]{david.ottaway@adelaide.edu.au}
%\thanks{}
%\affiliation{ Department of Physics and The Institute of Photonics and Advanced
%Sensing, The University of Adelaide, Adelaide, South Australia, Australia}


\date{\today}

\begin{abstract}
The physical degrees of freedom of a gravitational wave (GW) are imprints of the nature of gravity. We can test a gravity theory by searching for polarization modes beyond general relativity. The LIGO-Virgo collaboration analyzed several GW events in the O1 and O2 observing runs in the pure polarization framework, where they perform the Bayesian model selection between general relativity and the theory allowing only scalar or vector polarization modes. In this paper, we reanalyze the polarizations of GW170814 (binary black hole merger) and GW170817 (binary neutron star merger) in the improved framework of pure polarizations including the angular patterns of nontensorial radiation.
We find logarithms of the Bayes factors of 2.775 and 3.636 for GW170814 in favor of the pure tensor polarization against pure vector and scalar polarizations, respectively. These Bayes factors are consistent with the previous results by the LIGO-Virgo collaboration, though the estimated parameters of the binaries are significantly biased. For GW170817 with the priors on the location of the binary from NGC4993, we find logarithms of the Bayes factors of 21.078 and 44.544 in favor of the pure tensor polarization against pure vector and scalar polarizations, respectively. These more strongly support GR, epecially compared to the scalar polarization, than the previous results by the LIGO-Virgo collaboration due to the location prior. In addition, by utilizing the orientation information of the binary from a gamma-ray burst jet, we find logarithms of the Bayes factor of 51.043 and 60.271 in favor of the pure tensor polarization against pure vector and pure scalar polarization, much improved from those without the jet prior.
\end{abstract}

%\keywords{}

\maketitle


\section{Introduction}
The observation of gravitational waves (GWs) from compact binary coalescences made it possible to test gravity theories including general relativity (GR) in a stronger regime of gravity \cite{Abbott2016g, Abbott2016e, Abbott2017, Abbott2017b, Abbott2018, Abbott2019, Abbott2018b, Abbott2019b}. Search for physical degrees of freedom of a GW is an powerful approach to test gravity in a model-independent way because the properties of the polarization modes differ in each gravity theory~\cite{Hayama2013a, Nishizawa2009a,  Callister2017, Philippoz2018, Isi2015, Isi2017c, Chatziioannou2012, Isi2017a, Takeda2018, Hagihara2019, Takeda2019}. 

A GW in GR can have only two tensor polarization modes (plus, cross)~\cite{Misner1973, Will2005, Maggiore2007, Creighton2011}. However, at most six polarizations are allowed in a generic metric theory: two tensor modes (plus, cross), two vector modes (vector x, vector y), and two scalar modes (breathing, longitudinal) \cite{Eardley1973a, Eardley1973b, Will1993}. GWs have been studied transparently by the Newman-Penrose formalism \cite{Newman1962a, Alves2010a,Hyun:2018pgn}.
For example, GWs in modified gravity theories such as scalar-tensor theory \cite{Brans1961, Fujii2003} and f(R) gravity \cite{Buchdahl1970, DeFelice2010, Sotiriou2010, Nojiri2010, Nojiri2017} can have scalar polarization modes in addition to tensor modes \cite{Eardley1973a, Will1993, Will2005,  Hou2018, Katsuragawa2019}. In contrast, up to six polarization modes are possible \cite{Alves2010a} in bimetric gravity theory \cite{Visser1997, Hassan2011} while up to five polarization modes are possible \cite{DeRham2011} in massive gravity theory \cite{Rubakov2008, DeRham2010}.

Searches for polarization modes of GWs observed with three detectors, GW170814 (binary black hole merger)  and GW170817 (binary neutron star merger), have been conducted in \cite{Abbott2017, Abbott2017b, Abbott2019}. Therein, they assume pure polarization theories with the same waveforms as in GR and replace the standard tensor antenna pattern functions with those for scalar or vector polarization modes. They reported logarithms of the Bayes factors of 2.30 and 3.00 for GW170814 and 20.81 and 23.09 for GW170817 in favor of the pure tensor polarization against pure vector and scalar polarizations, respectively.

However, the waveforms of GWs for nontensorial modes should depend on not only specific modifications of gravity but also the geometrical parameters of a system such as binary inclination through the angular pattern of nontensorial radiation \cite{Chatziioannou2012, Takeda2018}. In this paper, we study how such an inclination-angle dependence of the waveforms biases the parameter estimation of the GWs from compact binary mergers. Moreover, we analyze GW170814 and GW170817 in the improved framework of pure polarizations with nontensorial inclination dependence, taking into account the nontensorial radiation patterns.

This paper is organized as follows. We review a polarization test of GWs in \Section{Polarization test} and show the inclination dependence of GW radiation in modified gravity in \Section{Angular patterns of gravitational-wave radiation}. In \Section{Basics of Bayesian inference}, we describe the basics of Bayesian inference. In \Section{Bias by inclination dependence}, we present how inclination dependence affects the parameter estimation of Bayesian inference. In \Section{Pure polarization theory analysis}, we give the results of the pure polarization analysis for GW170814 and GW170817. Finally we devote the last \Section{sec:Discussions-Conclusion} to the discussions and conclusion of the paper. Throughout the paper we use the natural units.

\section{Detector signal}

\label{Polarization test}
Since the metric theory of gravity allows four nontensorial polarization modes in addition to tensorial modes in general \cite{Eardley1973a, Eardley1973b, Will1993}, a metric perturbation denoting a GW can be written as
\begin{equation}
\label{gw}
h_{ab}(t,\hat{\Omega})=\sum_A h_{A}(t)e^{A}_{ab}(\hat{\Omega}).
\end{equation}
Polarization indices $A$ run over $+,\times, x, y, b, l$, corresponding to plus, cross, vector x, vector y, breathing, and longitudinal polarization modes, respectively. $h_{A}(t)$ are the components of the GW for each polarization mode and $\hat{\Omega}$ is the sky direction of a GW source.  $e^{A}_{ab}(\hat{\Omega})$ are polarization basis tensors, which are defined in \cite{Takeda2018}. 

 The detector signal of the $I$-th GW detector is expressed as \cite{Forward1978, Tobar1999, Will2005, Poisson2014, Will2018}
 \begin{equation}
 \label{detector_signal}
 h_I(t,\hat{\Omega})=d_{I}^{ab}h_{ab}(t,\hat{\Omega})= \sum_A F_I^{A}(\hat{\Omega})h_A(t),
 \end{equation} 
 where $d_I^{ab}$ is the $I$-th detector tensor defined by
  \begin{equation}
d_I^{ab}:=\frac{1}{2}(\hat{u}_{I}^{a}\otimes\hat{u}_{I}^{b}-\hat{v}_{I}^{a}\otimes\hat{v}_{I}^{b}),
  \end{equation}
where $\hat{u}_I, \hat{v}_I$ are unit vectors along the arms of the I-th interferometric detector. $ F_{I}^A$ is  called the antenna pattern functions of the I-th detector for polarization $A$ defined by
  \begin{equation}
  \label{antenna}
  F_I^{A}(\hat{\Omega}):=d_I^{ab}e^{A}_{ab}(\hat{\Omega}).
  \end{equation}
  
In general, two unit vectors $\hat{u}_I(t), \hat{v}_I(t)$ depend on time due to the Earth's rotation\cite{Takeda2019}, but the antenna pattern functions of the second-generation GW detectors such as  Advanced LIGO (aLIGO) \cite{Aasi2015}, Advanced Virgo (AdV) \cite{Acernese2015} and KAGRA \cite{Somiya2012, KAGRACollaboration2020, KAGRACollaboration2013} can be regarded as constants in time because a GW signal from typical compact binary merger in the observational band is short enough to ignore the Earth's rotation.
The nature of gravity can be probed by extracting each polarization mode in the detector signal, because the possible polarization modes depend on a specific theory of gravity.

\section{Angular patterns of gravitational-wave radiation}
\label{Angular patterns of gravitational-wave radiation}

A metric perturbation of a GW radiated from a source in modified gravity theories can be derived by the quadrupole formula without taking the transeverse-traceless projection,
 \begin{equation}
 h_{ab}(t, \bm{x})=\frac{2}{r}\ddot{M}_{ab}(t-r/c),
% h_{ab}(t, \bm{x})=\frac{1}{r}\frac{2G}{c^4}\ddot{M}_{ab}(t-r/c),
 \label{gw-quadrupole}
 \end{equation}
where $M_{ab}$ is the quadrupole moment of a mass distribution. $a, b$ run over the source coordinate $\{x_1, x_2, x_3\}$. Here we simply consider a circular motion of a binary in the $x_1$-$x_2$ plane.

\begin{figure}
\begin{center}
\includegraphics[width=\hsize]{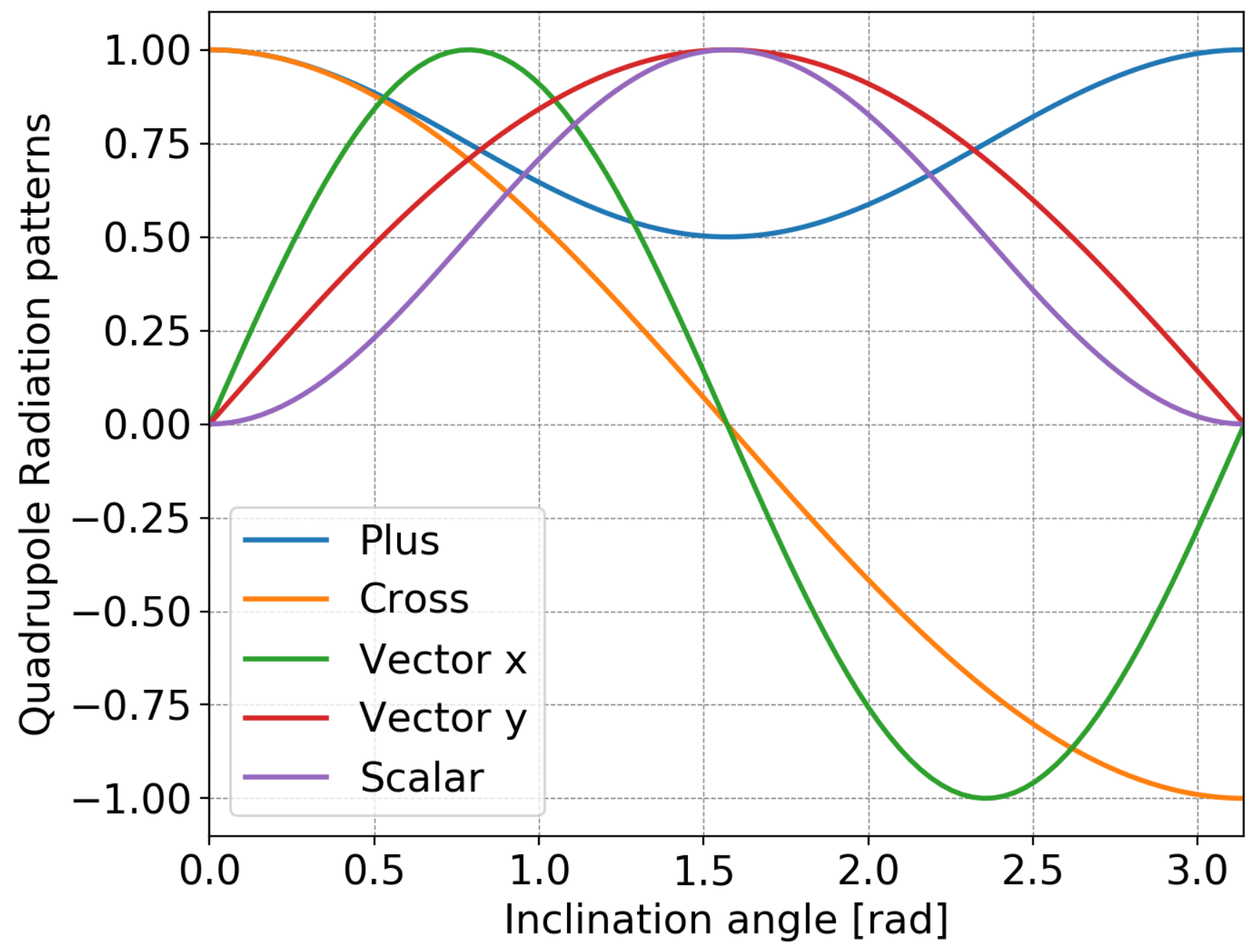}
\end{center}
\caption{Inclination-angle dependence of the GW quadrupole radiation for the tensorial and nontensorial polarization modes.}
\label{inclination_dependence_fig}
\end{figure}

In GR, only plus and cross tensor modes are kept after transverse-traceless projection \cite{Maggiore2007}. In modified gravity, additional degrees of freedom of the theory or the breaking of the gauge symmetries result in leading to additional non-transverse-traceless degrees of freedom for a GW.
%\smc{(If a graviton is massive, this statement is true. But if a tensor graviton is massless, the theory still keeps the gauge symmetry of GR because the GW equations of motion for the tensor and scalar modes are decoupled. The nontensorial mode can exist merely because the theory has a scalar degree of freedom in gravity sector.)} 
As a result, non-transverse-traceless components induce nontensorial polarization modes. According to \Eq{gw-quadrupole}, considering a circular binary system in the direction of $\hat{n}=(\sin{\iota}\cos{\phi}, \sin{\iota}\sin{\phi}, \cos{\iota})$, the amplitudes of the nontensorial polarization modes can be calculated as \cite{Takeda2018}
\begin{equation}
%h_x=-\frac{4G\mu\omega_{s}^2R^2}{rc^4}\frac{\sin{2\iota}}{2}\cos{(2\omega_{s}t_{\rm{ret}}+2\phi)},
h_x=-\frac{4\mu\omega_{s}^2R^2}{r}\frac{\sin{2\iota}}{2}\cos{(2\omega_{s}t_{\rm{ret}}+2\phi)},
\end{equation}
\begin{equation}
%h_y=-\frac{4G\mu\omega_{s}^2R^2}{rc^4}\sin{\iota}\sin{(2\omega_{s}t_{\rm{ret}}+2\phi)},
h_y=-\frac{4\mu\omega_{s}^2R^2}{r}\sin{\iota}\sin{(2\omega_{s}t_{\rm{ret}}+2\phi)},
\end{equation}
\begin{equation}
%h_b=-\frac{4G\mu\omega_{s}^2R^2}{rc^4}\frac{\sin^2{\iota}}{2}\cos{(2\omega_{s}t_{\rm{ret}}+2\phi)},
h_b=-\frac{4\mu\omega_{s}^2R^2}{r}\frac{\sin^2{\iota}}{2}\cos{(2\omega_{s}t_{\rm{ret}}+2\phi)},
\end{equation}
\begin{equation}
%h_l=\frac{4G\mu\omega_{s}^2R^2}{rc^4}\frac{\sin^2{\iota}}{\sqrt{2}}\cos{(2\omega_{s}t_{\rm{ret}}+2\phi)},
h_l=\frac{4\mu\omega_{s}^2R^2}{r}\frac{\sin^2{\iota}}{\sqrt{2}}\cos{(2\omega_{s}t_{\rm{ret}}+2\phi)},
\end{equation}
Here, $\omega_s$, $\mu$, $R$, $t_{\rm{ret}}$ are the orbital angular frequency of the binary stars, the reduced mass, and  the orbital radius, and the retarded time defined by $t_{\rm{ret}}=t-r/c$, respectively. \Fig{inclination_dependence_fig} shows the dependence of the GW amplitudes on the inclination angle, $\iota$.

The amplitude modification  and the phase evolution depend on the specific alternative theory of gravity.
However, the above inclination dependence is general among metric theories of gravity because they are determined by the geometry of the system. The inclination dependence for each polarization mode is encoded into a general framework of the inspiral waveforms, i.e., the parametrized-post Einsteinian framework \cite{Chatziioannou2012}.
%The parametrized post-Einsteinian waveform for quadrupole radiation with full polarization contents that is a framework including the modified response function beyond GR polarizations has been reported. In the framework, the waveform of gravitational waves from binary inspiral is given by \cite{Chatziioannou2012}

%\begin{equation}
%\begin{split}
%\tilde{h}_{\rm ppE, 2}(f)=&\tilde{h}^{\rm GR}_{\rm MD}(1+c\beta u^{b+5}_{2})e^{2i\beta u^{b}_{2}}e^{i\kappa u^{k}_{2}}+\{\alpha_b F^b\sin^2{\iota}\\
%\tilde{h}_{\rm ppE, 2}(f)=&\tilde{h}^{\rm GR}_{\rm MD}(1+\beta u^{b+5}_{2})e^{2i\beta u^{b}_{2}}e^{i\kappa u^{k}_{2}}+\{\alpha_b F^b\sin^2{\iota}\\
%&+\alpha_l F^l\sin^2{\iota}+\alpha_{x}F^{x}\sin{2\iota}+i\alpha_{y}F^{y}\sin{\iota}\}\\
%&\times\frac{\mathcal{M}}{d_L}u_{2}^{-7/2}e^{-i\Psi^{(2)}_{\rm GR}}e^{2i\beta u^{b}_{2}}e^{i\kappa u^{k}_{2}},
%\label{ppe}
%\end{split}
%\end{equation}
%with 
%\begin{equation}
%\begin{split}
%\tilde{h}^{\rm GR}_{\rm MD}=&-\{F^{+}(1+\cos^2{\iota})+2iF^{\times}\cos{\iota}\}\\
%&\times\left(\frac{5\pi}{96}\right)^{1/2}\frac{\mathcal{M}^2}{d_L}(\pi\mathcal{M}f)^{-7/6}e^{-i\Psi^{(2)}_{\rm GR}}.
%\end{split}
%\end{equation}

In the previous works \cite{Abbott2017b, Abbott2019}, they replace the standard tensor antenna pattern functions in the detector signal with those for scalar or vector polarization modes. In the method accompanied only by the replacement of the antenna pattern functions, it is assumed that the inclination dependences for the scalar and vector polarization modes are the same as that for the tensor modes. 
In other words, for example, the pure vector polarization model takes the following signal model,
\begin{align}
 h_I(t,\hat{\Omega}) &=F_I^{x}(\hat{\Omega})\frac{1+\cos^2{\iota}}{2}h_{+, {\rm GR}}(t) \nonumber \\
 &+F_I^{y}(\hat{\Omega})\cos{\iota}\,h_{\times, {\rm GR}}(t).
 \label{pure_vector_without_inc}
\end{align}
Here, $h_{+, {\rm GR}}(t)$ and $h_{\times, {\rm GR}}(t)$ are the waveforms for the plus and cross modes of the GW from a compact binary coalescence in GR without the factors of the inclination angle, respectively.

In this paper, we search for the pure polarization modes by replacing not only the antenna pattern functions for nontensorial modes but also the angular patterns of the nontensorial radiation, $\iota$ above. We adopt the following signal model,
\begin{equation}
 h_I(t,\hat{\Omega})=F_I^{x}(\hat{\Omega})\sin{2\iota}\,h_{+, {\rm GR}}(t)+F_I^{y}(\hat{\Omega})\sin{\iota}\,h_{\times, {\rm GR}}(t),
 \label{pure_vector_with_inc}
\end{equation}
for our pure vector polarization model and
\begin{equation}
 h_I(t,\hat{\Omega})=F_I^{b}(\hat{\Omega})\sin^2{\iota}\,h_{+, {\rm GR}}(t),
 \label{pure_scalar_with_inc}
\end{equation}
for our pure scalar polarization model. In these models, the dependences of the antenna pattern functions and inclination angle are derived from general consideration independent of a specific theory of modified gravity.

\section{Basics of Bayesian inference}
\label{Basics of Bayesian inference}

%According to Bayes theorem, the posterior distribution $p(\theta|d)$ is given by 
%\begin{equation}
%p(\theta|d)=\frac{\mathcal{L}(d|\theta)\pi(\theta)}{\mathcal{Z}},
%\end{equation}
%in the model M.
%Here, $\mathcal{L}$ is the likelihood function of the data $d$ given the parameters $\theta$, $\pi(\theta)$ is the prior distribution function for $\theta$, and $\mathcal{Z}$ is the evidence defined by
%\begin{equation}
%\mathcal{Z}:=\int d\theta \mathcal{L}(d|\theta)\pi(\theta).
%\end{equation}
%\smc{The likelihood function is a function to introduce a noise model. (This does not make sense to me. First assuming Gaussian noise, the likelihood function can have the Gaussian form.)}  For gravitational-wave astronomy, we basically assume a Gaussian-noise likelihood function,
%\begin{equation}
% \mathcal{L}(d|\theta)= \frac{1}{2\pi\sigma^2}\exp{\left(-\frac{1}{2}\frac{|d-\mu(\theta)|^2}{\sigma^2}\right)}.
%\end{equation}
%\smc{($2\pi\sigma^2 \longrightarrow {\sqrt{2\pi}\sigma}$?)}
%Here, $\mu(\theta)$ is a template for the gravitational-wave waveform given $\theta$ and $\sigma$ is the detector noise.
%\smc{(Is it better to use $h(t,\theta)$ explicitly instead of $\mu(\theta)$?)}

We analyze data under either of the 3 hypotheses: pure scalar hypothesis $\HS$, pure vector hypothesis $\HV$ and pure tensor hypothesis $\HT$, where
\begin{eqnarray}
\HS\text{: }h_I(t,\hat{\Omega})=&F_I^{b}&(\hat{\Omega}) h_b(t), \\
\HV\text{: }h_I(t,\hat{\Omega})=&F_I^{x}&(\hat{\Omega}) h_x(t) + F_I^{y}(\hat{\Omega}) h_y(t), \\
\HT\text{: }
h_I(t,\hat{\Omega})=&F_I^{+}&(\hat{\Omega})\frac{1+\cos^2{\iota}}{2}h_{+, {\rm GR}}(t) \nonumber\\
&+&F_I^{\times}(\hat{\Omega})\cos{\iota}\,h_{\times, {\rm GR}}(t).
\label{tensor}
\end{eqnarray}
Here, we assume that the waveforms for the nontensorial modes are those of the tensor modes except for the inclination dependence,
\begin{align}
h_b(t)&=\sin^2{\iota}\,h_{+, {\rm GR}}, \\
h_x(t)&=\sin{2\iota}\,h_{+, {\rm GR}}(t), \\ h_y(t)&=\sin{\iota}\,h_{\times, {\rm GR}}(t).
\end{align}
Since the antenna patterns of an interferometer for the breathing mode and the longitudinal mode have the same functional form and are degenerated, it is impossible to distinguish two scalar modes. Thus, the model includes only breathing mode under $\HS$.

We infer source parameters $\bm{\theta}$ under each hypothesis, where the parameters we consider in this work are
\begin{equation}
\bm{\theta}=(\alpha, \delta, \iota, \psi, d_L, \tc, \phic, m_1, m_2, \chi_1, \chi_2, \Lambda_1, \Lambda_2).
\end{equation}
$\alpha$ and $\delta$ represent the right ascension and declination of the binary.
$\psi$ is the polarization angle of GWs, and $\iota$ and $\psi$ determine the direction of the orbital angular momentum.
$d_L$ is the luminosity distance to the binary.
$\tc$ and $\phic$ is the time and phase at coalescence, respectively.
$m_1$ and $m_2$ are detector-frame masses of the primary and the secondary stars.
$\chi_1$ and $\chi_2$ are dimensionless spins of the primary and secondary stars.
$\Lambda_1$ and $\Lambda_2$ are tidal deformability parameters of the primary and secondary stars.

Our analysis is based on the Bayesian inference, where the posterior probability distribution is calculated through the Bayes' theorem,
\begin{equation}
p(\bm{\theta}| \dIs, \HX) = \frac{p(\bm{\theta}) p(\dIs | \bm{\theta}, \HX)}{p(\dIs|\HX)},
\end{equation}
where X is S, V or T.
$p(\bm{\theta})$ is referred to as a prior probability distribution, which encodes our knowledge or belief on the source parameters.
We apply the standard prior used by the LIGO-Virgo collaboration (See Appendix B in \cite{LIGOScientific:2018mvr}).
The prior range of $\chi_1$ and $\chi_2$ is $-0.99<\chi_1,\chi_2<0.99$.
The prior on $\Lambda_1$ and $\Lambda_2$ are $\delta(\Lambda_{1,2})$ for GW170814 based on the assumption that it is binary black hole and uniform over $\Lambda_1, \Lambda_2<5000$ for GW170817.
$p(\dIs | \bm{\theta}, \HX)$ is referred to as a likelihood function, which is determined by the properties of the instrumental noise. 
We apply the standard Gaussian noise likelihood, which is given by Eq.~(8) in \cite{LIGOScientific:2018mvr}.
The lower frequency cutoff for the likelihood calculations is $20\unit{Hz}$ for GW170814 and $23\unit{Hz}$ for GW170817, which are the same as those in the analysis of the LIGO-Virgo collaboration \cite{Abbott:2018wiz, LIGOScientific:2018mvr}.
To obtain the probability density functions of the source parameters we are interested in, we generate thousands of random samples following the posterior distribution and make their histograms.

%%%%% Mode detailed explanations on prior %%%%%
% We apply isotropic prior on $(\alpha, \delta)$.
% The prior on $(\iota, \psi)$ is determined so that the orbital angular momentum is distributed isotropically.
% The prior on $r$ is proportional to $r^2$ based on the assumption that the binary is distributed uniformly in comoving volume and cosmological corrections are negligible.
% We apply uniform prior for $\tc$, $\phic$, $m_1$ and $m_2$.
% The prior on $\chi_1$ and $\chi_2$ is aligned-z prior {\color{red}(CITEME)} with the spin range of $-0.99<\chi_1,\chi_2<0.99$.
% The prior on $\Lambda_1$ and $\Lambda_2$ are $\delta(0)$ for GW170814 based on the assumption that it is binary black hole and uniform over $\Lambda_1, \Lambda_2<5000$ for GW170817.

$p(\dIs|\HX)$ is referred to as evidence, which quantifies how much the hypothesis $\HX$ is favored by the observed data.
The model selection between the hypotheses $\HX$ and $\mathcal{H}_{\text{Y}}$ is done by calculating the Bayes factor defined by the ratio of these evidence,
\begin{equation}
B_{\text{XY}} := \frac{p(\dIs|\HX)}{p(\dIs|\mathcal{H}_{\text{Y}})}.
\end{equation}
One of our main goals is to calculate $B_{\text{TS}}$ and $B_{\text{TV}}$ to determine how much the pure tensor model, that is, GR is preferred compared to the pure scalar and vector models.

To generate random samples and calculate evidence, we utilize the Bilby software \cite{Ashton2019, Romero-Shaw:2020owr} and the cpnest sampler \cite{cpnest}, which is one of the implementations of the nested sampling technique \cite{Skilling:2006gxv}.
As a template waveform, we apply IMRPhenomD \cite{Khan2016} for GW170814 and IMRPhenomD\_NRTidal \cite{Dietrich:2018uni} for GW170817.
For GW170817, the generation of templates is computationally costly, which makes the parameter inference time-consuming and practically intractable.
To speed up the analysis, we applied the focused reduced order quadrature technique \cite{Morisaki:2020oqk}, where the reduced order quadrature basis vectors \cite{Canizares:2014fya, Smith:2016qas} of templates are constructed within a narrow range of the chirp mass.

\section{Parameter estimation Bias by inclination dependence}
\label{Bias by inclination dependence}

It may bias the parameter estimation to ignore the inclination dependence of scalar modes or vector modes, which was done in the LIGO-Virgo analysis.
Thus, we conduct the injection test to investigate such biases in a pure polarization theory.

We inject the GW170814-like signal whose waveform has the vector inclination dependence in \Eq{pure_vector_with_inc} and analyze the signal in the framework of the pure vector theory with the vector antenna pattern functions, but, with the tensor inclination dependence, \Eq{pure_vector_without_inc}. The difference between the injected signal and the search template is the inclination angle dependence.

\begin{figure}
\begin{center}
\includegraphics[width=\hsize]{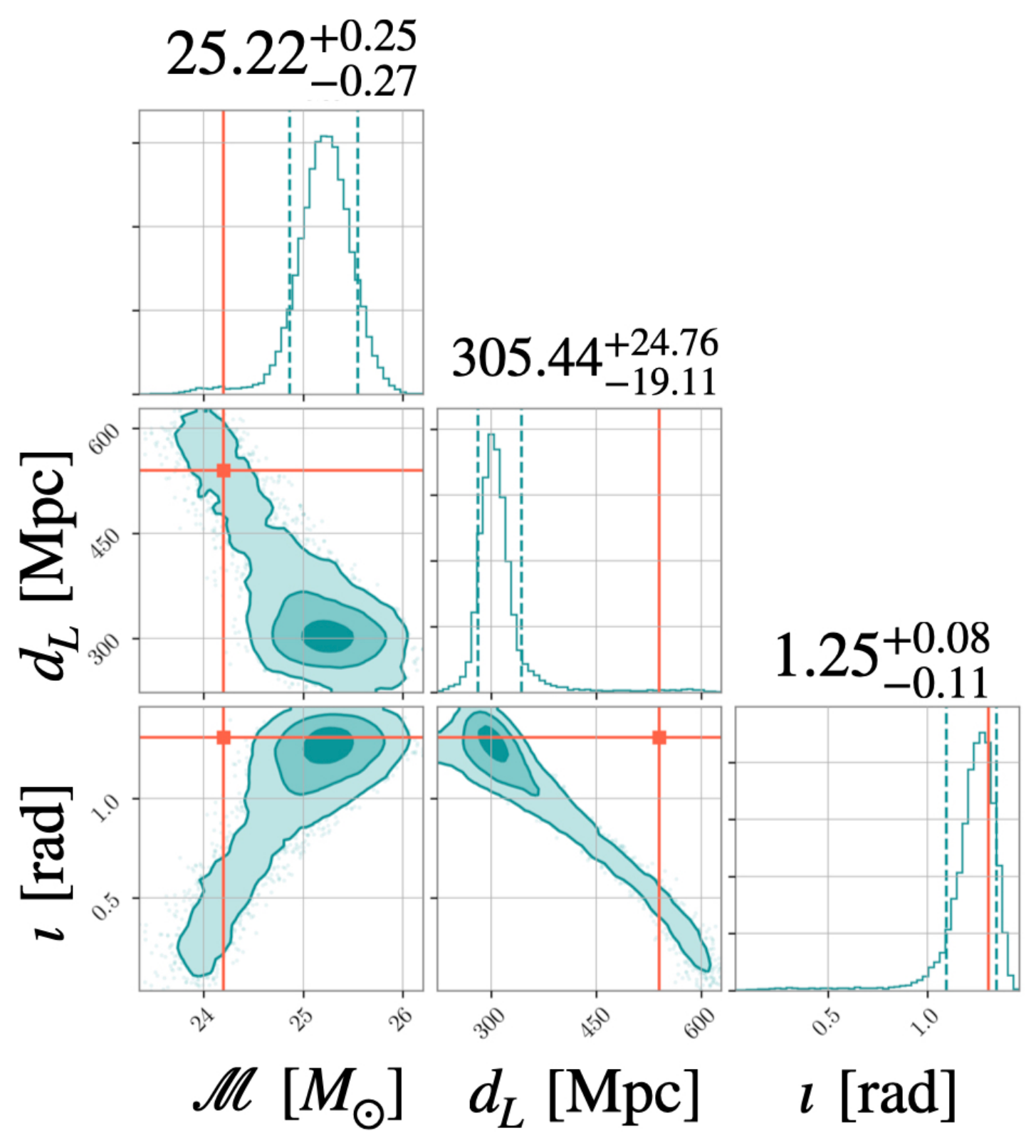}
\end{center}
\caption{Parameter estimation results of the signal injection in the pure vector theory. The posteriors for the chirp mass in the source frame, the luminosity distance, and the inclination angle are shown. The red lines show the injected values. The vertical dotted lines in the marginalized distributions show the $90\%$ confidence intervals. We inject the binary black hole signal of $\iota=75\unit{deg}= 1.31\unit{rad}$ with the vector radiation patterns but analyze the signal with tensor radiation pattern.}
\label{inject_131}
\end{figure}

%\done{(Just for curiosity, the solar mass, ${M_\odot}$ or ${{\rm M}_\odot}$, depends on a journal, e.g. ${M_\odot}$ for PRD and ApJ, ${{\rm M}_\odot}$ for PTEP.)}
\Fig{inject_131} shows the results of the parameter estimation in the case of a specific value of the inclination angle parameter. The injected parameters are the component masses in the source-frame, $m_1=30.5\ {M_\odot}$ and $m_2=25.3\ {M_\odot}$, the luminosity distance $d_L=540\unit{Mpc}$, the inclination angle $\iota=75\unit{deg}=1.31\unit{rad}$, the right ascension $\alpha=0.83\unit{rad}$, and the declination $\delta=-0.78\unit{rad}$. The results show when we use the waveform without appropriate inclination dependence in the Bayesian inference, it could produce the biases in the parameter estimation. When $\iota=75\unit{deg}$, a larger amount of vector GWs is radiated. However, if we adopt the waveform model with the tensor inclination dependence, the larger amplitude needs to be compensated by the parameters other than the inclination angle. As a result, the estimated luminosity distance becomes smaller than the injected value. The estimated chirp mass in the source frame is also shifted toward a larger value than the injected value.

 %When $\iota=0\unit{deg}$, the vector polarizations are not radiated. However, if we adopt the waveform model with tensor inclination dependence, the parameter estimation tends to show the results explaining nearly zero SNR with tensor inclination dependence, for example,  small chirp mass, large luminosity distance, and inclination angle $\sim\pi$. 

%\begin{figure}
%\begin{center}
%\includegraphics[width=\hsize]{pure_vector_inc_0_paper.pdf}
%\end{center}
%\caption{Parameter estimation results of the signal injection in pure vector theory. We inject the binary black hole signal of $\iota=0\unit{deg}$ with the vector radiation patterns, but we analyze the signal with tensor radiation pattern.}
%\label{pure_vector_inc_0}
%\end{figure}

%\begin{figure}
%\begin{center}
%\includegraphics[width=\hsize]{pure_vector_inc_45.pdf}
%\end{center}
%\caption{Parameter estimation results with cpnest in pure vector theory with inclination dependence. $\iota=45\unit{deg}$.}
%\label{pure_vector_inc_45}
%\end{figure}

%\clearpage

\section{Pure polarization test with real data}
\label{Pure polarization theory analysis}

In this section, we show our results of the analysis for GW170814 and GW170817 in the pure polarization framework.
From Gravitational Wave Open Science Center \cite{Abbott2019c}, we use the data of GW170814 whose duration is 4 seconds and sampling frequency is $4096\unit{Hz}$ and the data of GW170817 with the removal of glitch whose duration is 128 seconds and  sampling frequency is $4096\unit{Hz}$.

\subsection{GW170814}
\begin{figure}
\begin{center}
\includegraphics[width=\hsize]{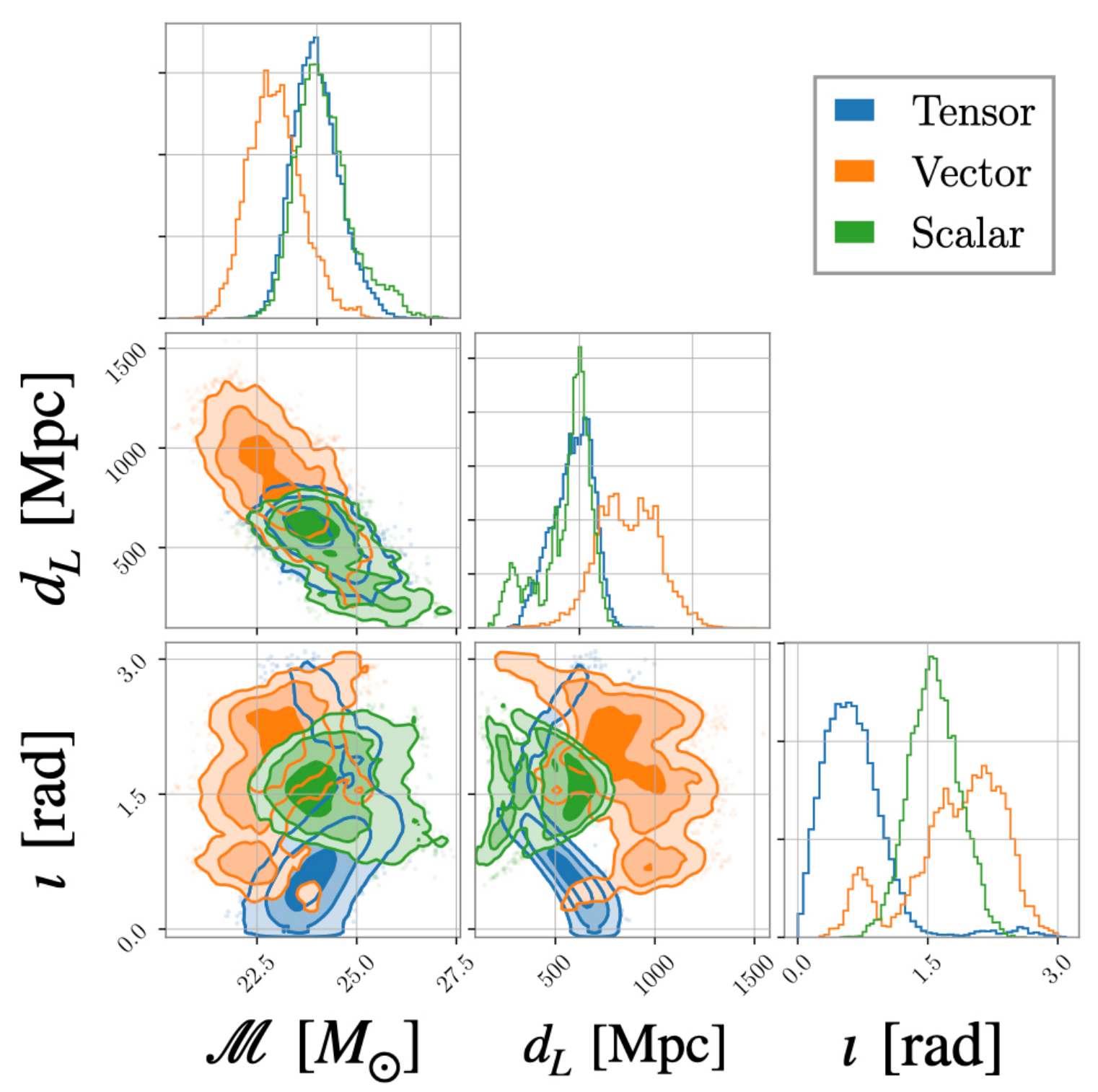}
\end{center}
\caption{The posterior distributions of GW170814 for the chirp mass, the luminosity distance, and the inclination angle in the pure polarization theories with the radiation patterns in the modified theories of gravity. The result of pure tensor theory (=GR) is shown in blue,  pure vector theory in orange, and pure scalar theory in green.}
\label{GW170814_chirp}
\end{figure}

\begin{figure}
\begin{center}
\includegraphics[width=\hsize]{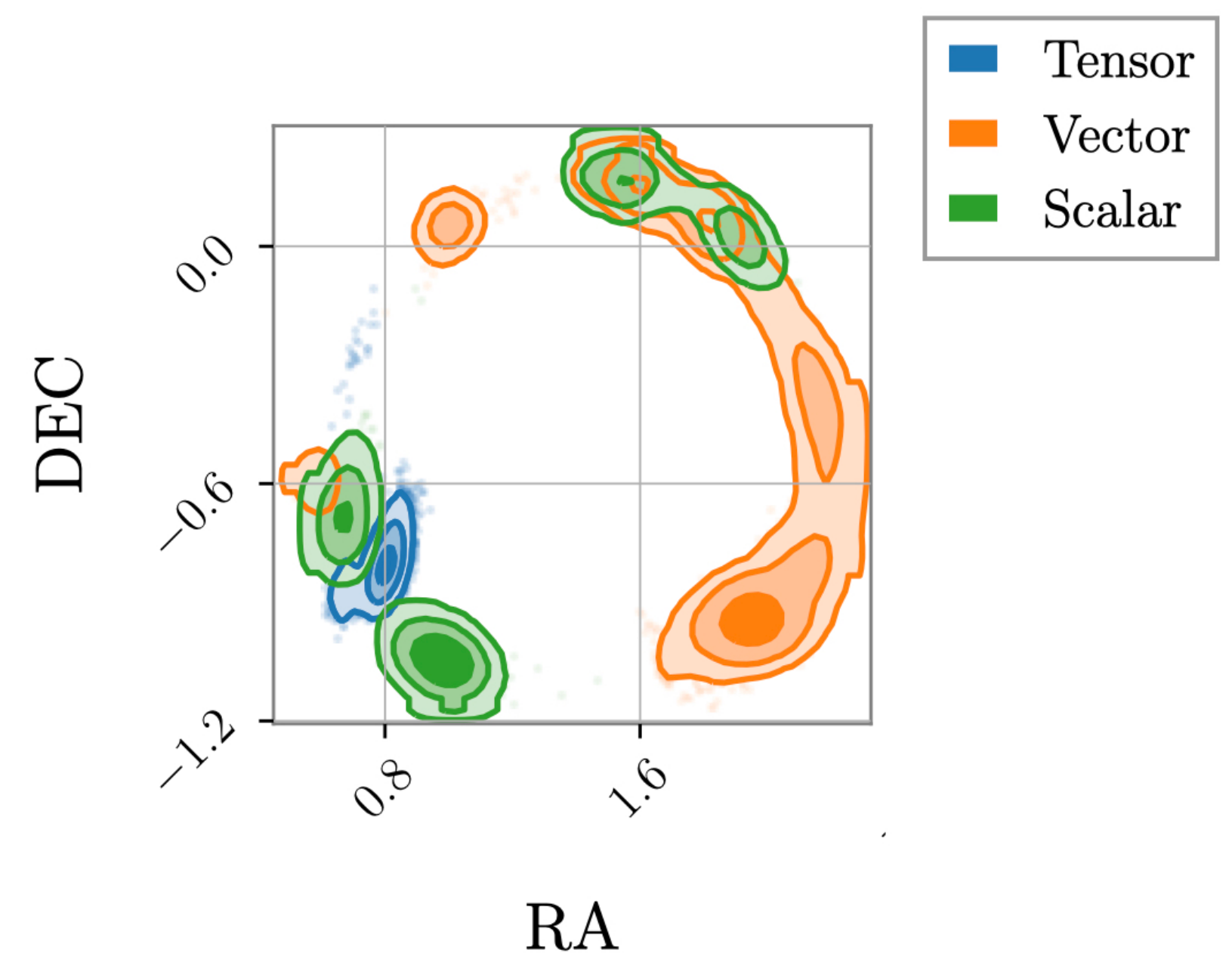}
\end{center}
\caption{The posterior distributions of GW170814 for the RA and the DEC in the pure polarization 
theories with the radiation patterns in the modified theories of gravity. The result of pure tensor theory (=GR) is shown in blue,  pure vector theory in orange, and pure scalar theory in green.}
\label{GW170814_ra_dec}
\end{figure}

GW170814 is a GW signal from a binary black hole merger observed by three detectors \cite{Abbott2017b}.
We perform the parameter estimation of GW170814 under $\HS$, $\HV$, and $\HT$. The results are shown in Figs.~\ref{GW170814_chirp} and \ref{GW170814_ra_dec}. The posterior probability distributions for the chirp mass in the source frame, the luminosity distance, and the inclination angle are shown in \Fig{GW170814_chirp} and those for the right ascension (RA) and the declination (DEC) are shown in \Fig{GW170814_ra_dec}. The results are shown in blue for $\HT$ (GR), in orange for $\HV$, and in green for $\HS$.

In \Fig{GW170814_chirp}, the estimated inclination angles have different values in the pure polarization theories, reflecting the quadrupole radiation pattern in each pure polarization theory. 
The estimated value of the luminosity distance under $\HV$ is slightly larger than those under $\HT$ and $\HS$. This is because of the correlation between the luminosity distance and the inclination angle. The luminosity distance and the inclination angle compensate each other in the parameter region where the correlation is strong under the hypothesis. As a result, the chirp mass in the source frame becomes smaller, corresponding to the  slightly larger luminosity distance. The RA and DEC are also adjusted through the antenna pattern functions to compensate the shift in other parameters and fit to the amplitude of the signal.

The logarightms of the Bayes factors are $\ln B_{TS}=3.636$ and $\ln B_{TV}=2.775$, which support the pure tensor hypothesis.

\subsection{GW170817}
GW170817 is a GW signal from a binary neutron star merger observed by three detectors \cite{Abbott2017}.
Also the gamma-ray burst, GRB 170817A, was observed independently by the Fermi Gamma-ray Burst Monitor \cite{Meegan2009}, and the Anti-Coincidence Shield for the Spectrometer for the International Gamma-Ray Astrophysics Laboratory \cite{Kienlin2003}. It is confirmed with high statistical significance that the GRB 170817A is associated with GW170817 \cite{Abbott2017c}. Furthermore, an optical \cite{Coulter2017} and near-infrared \cite{Tanvir2017} electromagnetic counterpart was localized to the position deviated by sub-arcseconds from the nucleus of the galaxy NGC 4993 half a day after the event~\cite{Abbott2017c}. We utilize these information about the location and orientation of the binary system. Here we estimate the parameters of GW170817 in the same way as the above analysis of GW170814. However, we impose the priors on the luminosity distance, the right ascension, and the declination of GW170817 from the host galaxy, NGC4993. The prior of the luminosity distance is the Gaussian distribution with the mean of 42.9 Mpc and the standard deviation of 3.2 Mpc. The right ascension and the declination are fixed to ${\rm RA} = 13{\rm h}09{\rm m}48{\rm s}.085$ and ${\rm DEC}=-23^{\circ}22'53''.343$ \cite{Abbott2017c}. From the estimation of the jet based on hydrodynamics simulations, the orientation of the binary system was constrained by $0.25\ {\rm rad}< \theta_{\rm obs} (d_L/41\ {\rm Mpc})< 0.45\ {\rm rad}$ \cite{Mooley2018, Hotokezaka2019}. Here, $\theta_{\rm obs}$ is the viewing angle and can be identified with the inclination angle, $\theta_{\rm obs}=\iota$ or $\theta_{\rm obs}=\pi-\iota$, under the assumption that the jet is perpendicular to the binary’s orbital plane. In the case of GW170817, we adopt $\theta_{\rm obs}=\pi-\iota$ from the estimated inclination angle. From our prior of the luminosity distance, we set the prior on the inclination angle in the range of $2.68\,{\rm rad} < \iota <2.92\, {\rm rad}$ optionally.  We call this prior the jet prior in the later analysis.

\subsubsection{Without prior of inclination}
\begin{figure}
\begin{center}
\includegraphics[width=\hsize]{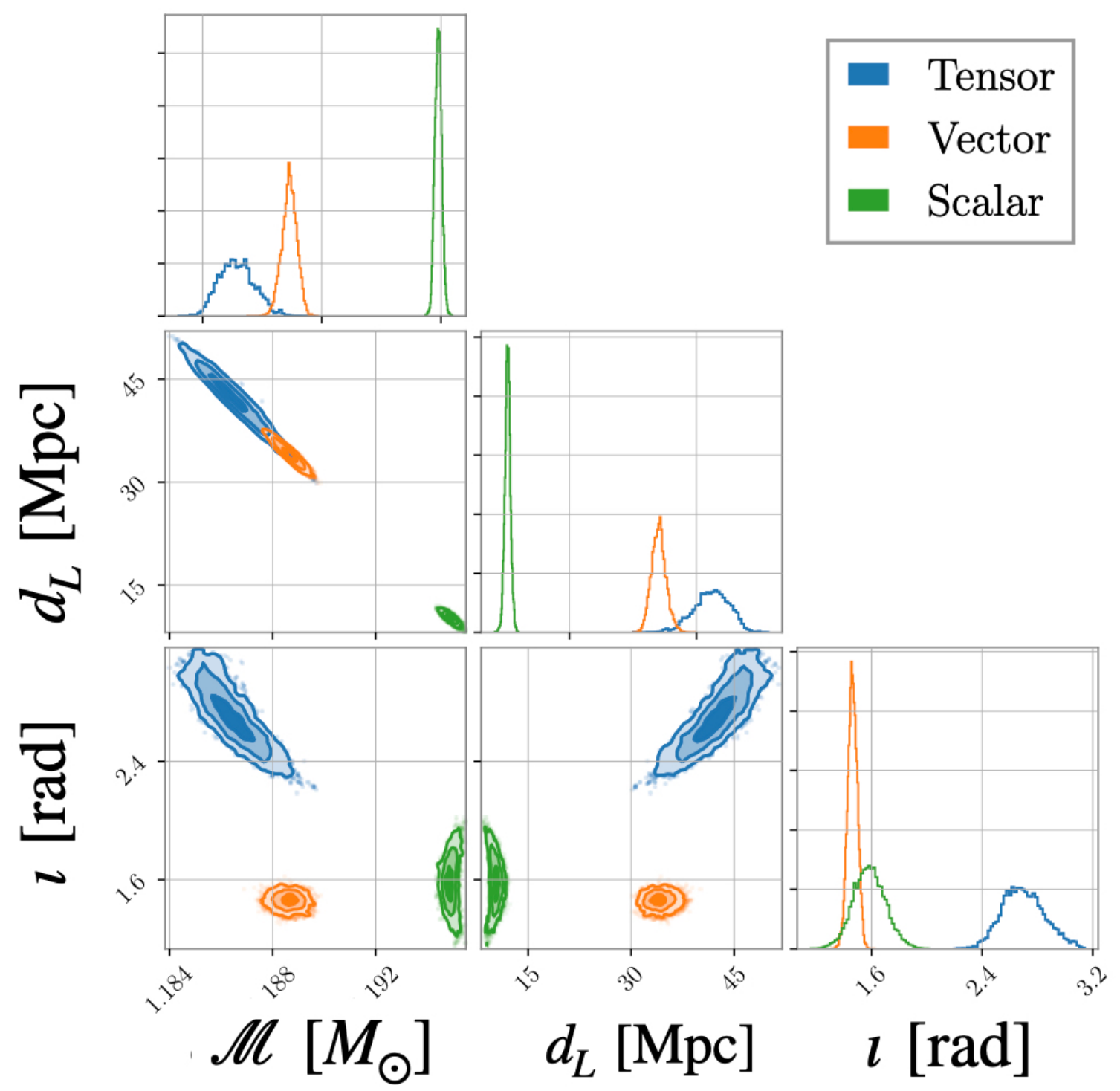}
\end{center}
\caption{The posterior distributions of GW170817 for the chirp mass, the luminosity distance, and the inclination angle in the pure polarization theories with the radiation patterns in the modified theories of gravity.  The result of the pure tensor theory(=GR) is shown in blue,  the pure vector theory in orange,  and the pure scalar theory in green. Here, we impose the prior on the RA, DEC, and the luminosity distance from NGC4993 but without the jet prior.}
\label{GW170817_inc}
\end{figure}

Here, we analyze GW170817 without the jet prior on the inclination angle.
\Fig{GW170817_inc} shows the result of the parameter estimation in each pure polarization hypothesis: the posterior probability distributions for the chirp mass in the source frame, the luminosity distance, and the inclination angle.
 %A base $e$ logarithm of the noise evidence is  -724566.347 and a logarithm of  the evidence is -724061.514. A logarithm of the Bayes factor is 504.834.
%A logarithm of the noise evidence is  -314675.166 and a logarithm of  the evidence is -314455.920.
%A logarithm of the Bayes factor is 219.247.
Again the results are shown in blue for $\HT$ (GR), in orange for $\HV$, and in green for $\HS$.
In comparison with GR, the amplitude parameters of the pure vector polarization are well determined. This is because the polarization angle and the phase at the coalescence time are degenerated in GR when the inclination angle is in the range of a nearly face-on. However, in the case of the vector polarization modes, the inclination angle is estimated to the  range of a nearly edge-on binary, reflecting the radiation pattern. Then it breaks the degeneracy between the polarization angle and the phase at the coalescence time.
%The log evidence is -314,476.998 and the log Bayes factor is 198.169.
%A logarithm of the Bayes factor between tensor and vector is 21.078. 

In the case of the scalar hypothesis $\HS$, the GW does not depend on the polarization angle due to its symmetry under the rotation around the propagation axis. Because of the breaking of the degeneracy between the polarization angle and the coalescence phase in GR, the coalescence phase is well determined and then the amplitude parameters are also well determined, compared to the GR case. The reason why the luminosity distance in the pure scalar model is estimated to be  significantly small is due to the values of the antenna pattern functions. The value of the antenna pattern function for the scalar mode is about 2-6 times smaller than those of the tensor and vector modes for the given direction of GW170817. To compensate the smallness of the antenna pattern function and fit to the amplitude of the signal, the luminosity distance needs to be small.
%The log evidence is -724164.070 and the Bayes factor is 402.278. A base $e$ logarithm of the Bayes factor between tensor and scalar is 102.566.
%The log evidence is -314500.460 and the log Bayes factor is 174.707.
%A logarithm of the Bayes factor between tensor and scalar is 44.544.

Our pure polarization search give the logarightms of the Bayes factors, $\ln B_{TS}=44.544$ and $\ln B_{TV}=21.078$ for GW170817, which more strongly supports the pure tensor hypothesis, especially compared to the pure scalar polarization hypothesis.

\subsubsection{With prior of inclination}
\begin{figure}
\begin{center}
\includegraphics[width=\hsize]{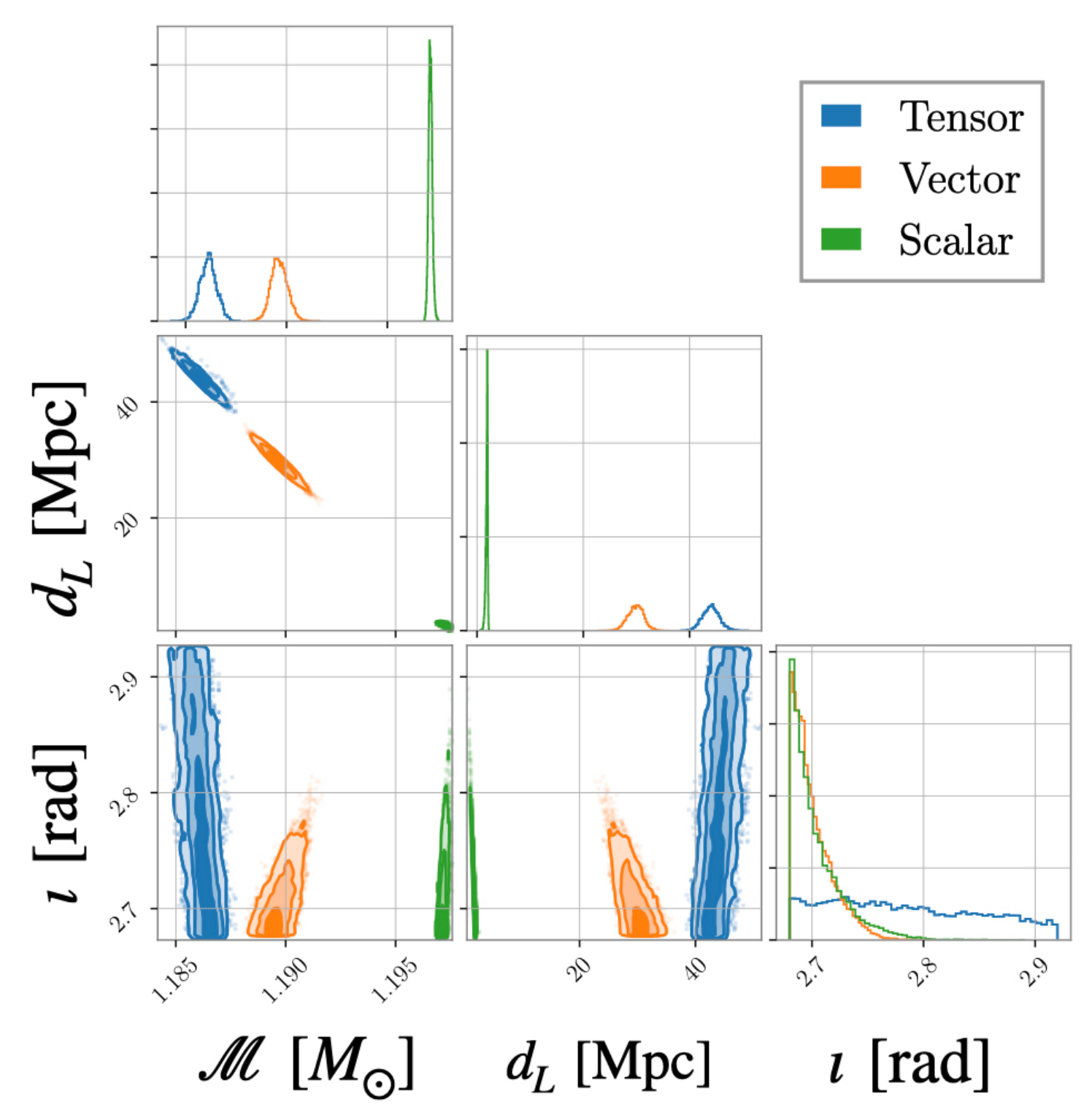}
\end{center}
\caption{The posterior distributions of GW170817 for the chirp mass, the luminosity distance, and the inclination angle in the pure polarization theories with the radiation patterns in the modified theories of gravity and the jet prior.  The result of the pure tensor theory(=GR) is shown in blue, the pure vector theory in orange, and the pure scalar theory in green. Here, we impose the prior on the RA, DEC, and the luminosity distance from NGC4993, and the jet prior from GRB170817A.}
\label{GW170817_inc_jet}
\end{figure}

Next, we analyze GW170817  with the jet prior on the inclination angle. 
%A base $e$ logarithm of the noise evidence is  -724566.347 and a logarithm of  the evidence is -724058.729. A logarithm of the Bayes factor is 507.618.
%A logarithm of the noise evidence is  -314675.166 and a logarithm of  the evidence is -314454.711.
 %a logarithm of the Bayes factor is 220.455. 
\Fig{GW170817_inc_jet} shows the result of the parameter estimation in each pure polarization hypothesis: the posterior probability distributions for the chirp mass in the source frame, the luminosity distance, and the inclination angle. Again the results are shown in blue for $\HT$ (GR), in orange for $\HV$, and in green for $\HS$. 

Since we consider the limited range of the inclination angle by the jet prior, the inclination angle is estimated near the lower bound under $\HS$ and $\HV$.
In comparison with \Fig{GW170817_inc}, although the values of the estimated luminosity distance slightly change accordingly, we obtain the similar trend in \Fig{GW170817_inc_jet}. 

The logarightms of the Bayes factors are $\ln B_{TS}=60.271$ and $\ln B_{TV}=51.043$, which strongly support the pure tensor hypothesis.

As in the case of GW170814, the distributions of the inclination angle in \Fig{GW170817_inc} are different each other in the pure theories, reflecting the radiation patterns.
However, for a binary neutron star event with an electromagnetic counterpart, we can utilize the prior distributions of location and inclination angle using the information about a host galaxy and a jet.
The fact that we observed the jet from the binary neutron stars means the binary system should be nearly face-on.
\Fig{inclination_dependence_fig} shows that the nontensorial polarization modes hardly radiate from such a nearly face-on binary.
Thus, binary neutron star events with jets can be utilized to distinguish the pure polarization theories.
As a result, we can obtain the larger Bayes factor or the stronger constraints of the pure polarization components.
 
%The log evidence is -724176.260 and the Bayes factor is 390.087. 
%A base $e$ logarithm of the Bayes factor between tensor and vector is 117.531. 
%The log evidence is -314,505.754 and the log Bayes factor is 169.413.

%The  log  evidence  is -724197.508  and  the  Bayes  factor  is  368.839. A  base e logarithm of the Bayes factor between tensor and scalar is 138.779.
%The  log  evidence  is -314514.982  and  the  log Bayes  factor  is  160.185.

\begin{figure}
\begin{center}
\includegraphics[width=\hsize]{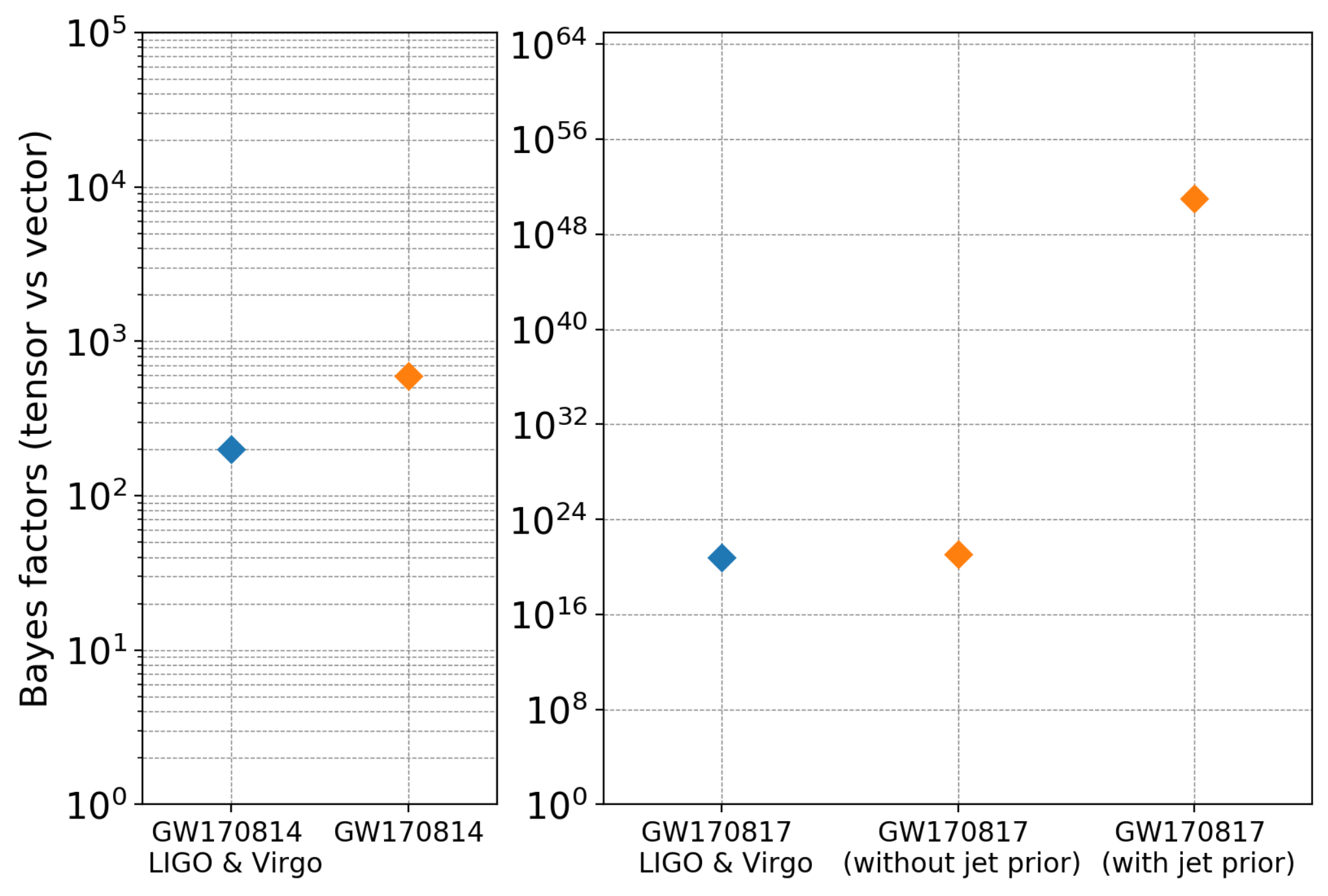}
\end{center}
\caption{The Bayes factors between the pure tensor polarization hypothesis and the pure vector polarization hypothesis are shown. The results by LIGO and Virgo are referred from \cite{Abbott2017b} for GW170814 and from \cite{Abbott2019} for GW170817 in blue.}
\label{result_vector}
\end{figure}

\begin{figure}
\begin{center}
\includegraphics[width=\hsize]{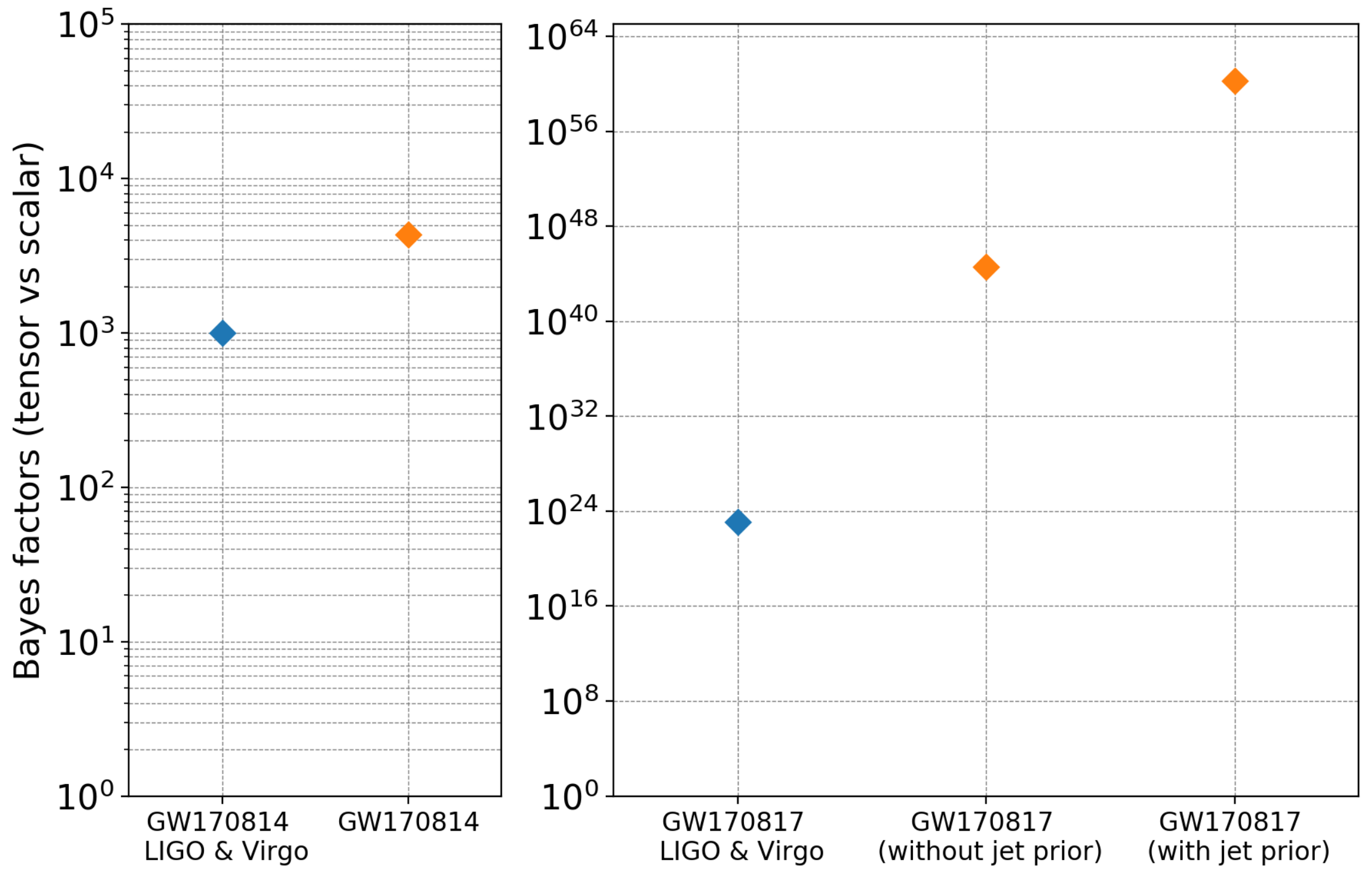}
\end{center}
\caption{The Bayes factors between the pure tensor polarization hypothesis and the pure scalar polarization hypothesis are shown. The results by LIGO and Virgo are referred from \cite{Abbott2017b} for GW170814 and from \cite{Abbott2019} for GW170817 in blue.}
\label{result_scalar}
\end{figure}

\section{Discussions and conclusion}
\label{sec:Discussions-Conclusion}

We studied how the radiation patterns or the inclination dependence of the nontensorial polarization modes affect the parameter estimation by Bayesian inference. We found that the values of the estimated amplitude parameters may be different from the true values if we adopt wrong radiation patterns. In addition, we conducted pure polarization tests of GW170814 (binary black hole merger) and GW170817 (binary neutron star  merger) under the three pure polarization hypotheses with nontensorial radiation patterns allowing only scalar, vector or tensor polarization modes. 
Figures~\ref{result_vector} and \ref{result_scalar} summarize our results of the Bayes factors for GW170814 and GW170817 between the scalar or vector hypotheses and the tensor hypothesis. For GW170814, we obtained the logarithmic Bayes factors of 2.775 and 3.636 in favor of the pure tensor polarization against the pure vector and scalar polarizations, respectively. These Bayes factors are consistent with the previous results by the LIGO-Virgo collaboration, though the estimated parameters of the binaries are significantly biased. In the analysis of GW170817, we utilized the information of the location and the orientation of the binary system from the electromagnetic counterpart. Especially, when a GW from binary neutron stars is observed with a jet, the binary system should be nearly face-on, in which the difference between the tensorial and nontensorial radiation patterns is large and helps distinguish the pure polarization states. For GW170817 with the known location of the electromagnetic counterpart, we found a logarithm of the Bayes factors of 21.078 and 44.544 in favor of the pure tensor polarization against pure vector and scalar polarization, respectively. Further imposing the prior on the observing angle of the GRB jet, they are improved to 51.043 and 60.271 respectively. These Bayes factors with the priors from the host galaxy and jet are much improved compared to the previous results by the LIGO-Virgo collaboration.

On the other hand, almost all theories of gravity predict the mixture of the polarization modes, for example, tensor and scalar modes.
A nearly face-on binary with a jet observed can bring us the information about the location and orientation of a binary in advance, but the amplitudes of nontensorial modes are expected to be relatively small. Therefore, an edge-on binary would also play an important role in searching for the mixture of polarizations. A study involving mixed-polarization modes is currently under way.

\section*{Acknowledgements}
H. T. acknowledges financial support received from the Advanced Leading Graduate Course for Photon Science (ALPS) program at the University of Tokyo. H.T. is supported by JSPS KAKENHI Grant No. 18J21016. S. M. is supported by JSPS KAKENHI Grant No. 19J13840. A.~N. is supported by JSPS KAKENHI Grant Nos. JP19H01894 and JP20H04726 and by Research Grants from Inamori Foundation.

\newpage


\bibliographystyle{h-physrev3}

\end{document}